\documentclass[aps,prb,preprint]{revtex4}
\usepackage{graphicx}
\usepackage{amsmath}
\begin{document}
\title{The role of van der Waals forces in water adsorption on metals}
\author{Javier Carrasco$^{1,a)}$, Ji\v{r}\'{i} Klime\v{s} $^{2}$, and Angelos Michaelides$^{3}$}
\affiliation{
$^1$ Instituto de Cat\'{a}lisis y Petroleoqu\'{i}mica, CSIC, Marie Curie 2, E-28049 Madrid, Spain\\
$^2$ Faculty of Physics and Center for Computational Nanoscience, University of Vienna,
Sensengasse 8/12, A-1090 Wien, Austria\\
$^3$  Thomas Young Centre, London Centre for Nanotechnology and
Department of Chemistry, University College London, London WC1E 6BT, UK\\
$^{a)}$ Electronic mail: j.carrasco@csic.es
}

\begin{abstract}
The interaction of water molecules with metal surfaces is typically weak and as a result van der Waals 
(vdW) forces can be expected to be of importance. 
Here we account for the 
systematic poor treatment of vdW forces in most popular density functional
theory (DFT) exchange-correlation functionals by applying accurate non-local
vdW density functionals.
We have computed the adsorption of a variety of 
exemplar systems including 
water 
monomer adsorption on Al(111), Cu(111), Cu(110),
Ru(0001), Rh(111), Pd(111), Ag(111), Pt(111), and 
unreconstructed Au(111), and small clusters (up to 6 waters) on Cu(110).
We show that non-local correlations contribute substantially to the water-metal
bond in all systems, whilst water-water bonding is much less affected by non-local 
correlations. 
Interestingly non-local correlations contribute more to the adsorption of water on
the reactive transition metal substrates than they do on the noble metals.
The relative stability, adsorption sites, and adsorption geometries
of competing  water adstructures rarely differ when comparing results obtained with
semi-local functionals and the non-local vdW density functionals, which
explains the previous success of semi-local functionals in characterizing adsorbed
water structures on a number of metal surfaces.
\end{abstract}

\maketitle

\section{Introduction}

An accurate atomistic description of the water-solid interface is crucial for
understanding many natural and technological processes
such as atmospheric ice formation or fuel cell reactions.
It is not surprising, therefore, that there has been an enormous amount
of research devoted to this task
in the past few decades \cite{thiel87,henderson02,hodgson09}. In recent
years, scanning probe 
techniques, in particular scanning tunneling microscopy (STM), have contributed
significantly to the field by providing detailed insight into the
structure and dynamics of water adstructures on the nanoscale. 
Generally such studies are, however, limited to
well-defined, single crystal metal surfaces at low temperature
and under ultra high vacuum (UHV) conditions \cite{hodgson09,carrasco12}.
As a result the water-metal interface has become
the workhorse system for understanding the basic chemistry and physics
of how water interacts with solid substrates in general.

Computer simulation techniques, in particular density functional theory (DFT),
have also played a central role in understanding the structure 
of water on metal surfaces \cite{hodgson09,carrasco12}. 
DFT has, for example, been 
essential in unravelling the structure of water overlayers on
Pd(111) \cite{cerda04}, Cu(110) \cite{carrasco09},
and Pt(111) \cite{nie10}.
Notwithstanding the clear value of DFT in this area,
the accuracy of the generalized gradient
approximation (GGA) functionals, which are generally used in such studies, remains an important open question.
This is mainly because the GGAs used fail to describe non-local van der Waals (vdW) dispersion forces, forces
which are expected to be relevant in water adsorption and weak adsorption systems in general \cite{hodgson09,feibelman08,feibelman09,carrasco12}.
This casts a shadow over all standard density functional studies of water 
on metals and hampers 
progress in the field as it puts accurate predictions about interesting phenomena such
as rates of heterogeneous ice nucleation beyond the reach of {\it ab initio}
methods at present.

Recent years have seen a number of exciting developments with DFT based
schemes for dealing with vdW dispersion forces (see e.g. refs
\cite{santra08,wu02,grimme04,jurecka07,dion04,becke05,silvestrelli08,lilienfeld04,tkatchenko09,klimes10,schimka10,vydrov10,klimes11} and for a recent review \cite{klimes12}). 
With this has come a flurry of interest in understanding the role of vdW forces in water adsorption systems. 
In particular, recent dispersion-corrected DFT studies have been reported for water adsorption on Rh(111) \cite{hamada10}, Cu(110) \cite{kumagai11}, Pd(111) \cite{poissier11,tonigold12},
Ag(111)\cite{tonigold12}, Au(111)\cite{tonigold12,nadler12}, and Pt(111)\cite{tonigold12} and by us on
Cu(110) and Ru(0001) \cite{carrasco11}. 
These studies
have indeed indicated that vdW dispersion forces should be accounted
for to properly describe the interaction of water molecules with the metal 
substrate underneath. In particular, it was shown that vdW dispersion forces
explain a long standing
discrepancy with respect to the relative thermodynamic stability of bulk ice and
wetting layers on Cu and Ru surfaces \cite{carrasco11}.
However, despite this work, important issues remain poorly understood, such as the relative importance
of vdW forces on different metal surfaces. Another open question is the role of vdW in determining the most
stable adsorption site and adsorption structures for hydrogen bonded water clusters on metals. 

Here we tackle these and related questions by exploring
in detail the role of vdW dispersion forces in the 
adsorption of water monomers and clusters on a number of metal surfaces.
The particular approach we use to treat vdW is the
non-local van der Waals density functional (vdW-DF) of Langreth and
Lundqvist and co-workers \cite{dion04} and some of its offspring \cite{klimes10}.
These functionals from the vdW-DF family have shown great potential when
applied to a number of systems where dispersion forces are important
\cite{langreth09,klimes10,klimes11,kelkkanen11,zhang11,santra11,hu11,carrasco11,addato11,lawton11,lew11,silvestrelli12,li12,lawton12,fernandez-torre12,forster12}.
From the current study we find that non-local correlations contribute substantially to the adsorption
of water monomers
and clusters on metal surfaces in general.
The contribution to the binding from non-local correlations varies from substrate to substrate and is actually greater
on the more reactive transition metal surfaces than on the noble metals.
Although vdW enhances the bonding with the substrate, the water-water interaction energies 
within the clusters remains largely unaffected by the inclusion of vdW dispersion 
forces.
In addition the geometries of the water monomers and clusters on the 
various surfaces are not affected to any great extent by the inclusion 
of vdW dispersion forces, which explains the previous success \cite{hodgson09,carrasco12}
of GGA functionals in predicting adsorption
structures of water on metals.

\section{Methodology and computational details}\label{sec:met}

Density functional theory and supercell periodic models were used within
the VASP 5.2 code \cite{vasp93,vasp96}. Total energies and electron densities were
computed with various different exchange-correlation functionals: (i) the
semi-local Perdew-Burke-Ernzerhof (PBE) \cite{pbe}; (ii) the non-local vdW-DF
of Dion {\it et al.} \cite{dion04}; and (iii) a couple of modified
versions of the original vdW-DF, where the original
GGA exchange
functional has been replaced by an optimized PBE (optPBE) or optimized Becke88
(optB88) in order to improve the accuracy of the vdW-DF scheme  \cite{klimes10}.
All vdW-DF calculations were carried out self-consistently within VASP as 
implemented by Klime\v{s} {\it et al.} \cite{klimes11} using the approach of
Rom\'{a}n-P\'{e}rez and Soler \cite{roman-perez09}.
In vdW-DF the exchange-correlation energy is calculated by adding three
different terms: a GGA exchange energy, the local 
correlation energy obtained within the local density approximation (LDA), 
and a non-local correlation energy based on electron densities interacting via a model
response function. In the original vdW-DF the GGA exchange is 
obtained with the revPBE functional \cite{revpbe}. However in ref. [\cite{klimes10}]
alternative exchange functionals to revPBE were developed which can improve
significantly the accuracy of the vdW-DF method.
In all calculations the core electrons were
replaced by projector augmented wave (PAW) potentials \cite{paw}, whilst
the wavefunctions of the valence electrons were expanded in plane-waves with a cut-off
energy of 600 eV. PBE-based PAW potentials were used for all calculations.
A Monkhorst-Pack \cite{monkhorst} grid with 12$\times$12$\times$1
{\bf k}-point sampling per (1$\times$1) unit cell was used.

In the case of water monomer adsorption, the close-packed (111) and (0001)
surfaces were modeled by $p$(2$\times$2) unit cells, containing
6 atomic layers separated by at least 14 \AA\ of vacuum (23 \AA\ when computing
binding curves). In the case of adsorbed water
dimers on Cu(110) we considered a $p$(3$\times$5) unit cell
containing 4 atomic layers, whilst in the cases of adsorbed 
trimers, tetramers, pentamers, and hexamers a $p$(4$\times$6) unit cell was employed. 
The metal atoms in the 3 (2 in the case of adsorbed water cluster models)
bottom layers were fixed at their bulk-truncated positions
during structure optimization procedures. We note that PBE lattice constants were
used for all functionals. This is likely to influence very little the adsorption energies
obtained with vdW functionals, since, for example, lattice constants are typically
within 0.010 of PBE for optB88-vdW \cite{klimes11}. In all cases a
dipole correction along the direction perpendicular to the metal surface was applied
\cite{neugebauer92,makov95} and geometry optimizations were performed with a
residual force threshold of 0.015 eV/\AA.

Adsorption energies per water molecule were computed as follows:

\begin{equation}\label{adene}
	E_{\rm ads} = \frac{E[{\rm H_2O/M}] - E[{\rm M}]- nE[{\rm H_2O}]}{n},
\end{equation}

\noindent where $E[{\rm H_2O/M}]$ is the total energy
of the adsorbed $n$ H$_2$O molecules, $E[{\rm M}]$ is the total energy of the
relaxed bare metal slab and $E[{\rm H_2O}]$ is the total energy of an 
isolated gas phase H$_2$O molecule. Within this definition a negative adsorption 
energy corresponds to a favorable (exothermic) adsorption process.

In order to analyze the role of van der Waals dispersion forces on adsorption, we have
decomposed $E_{\rm ads}$
in to different energy contributions \cite{michaelides04}. We define the
water-water contribution, $E_{\rm gas}^{\rm ww}$, to $E_{\rm ads}$ as:

\begin{equation}\label{eq:ww}
 E_{\rm gas}^{\rm ww} = \frac{E^{\rm tot}[{n \rm H_2O}] - nE^{\rm tot}[{\rm H_2O}]}{n},
\end{equation}

\noindent where $E^{\rm tot}[{n \rm H_2O}]$ is the total energy of the water structure in the absence of 
the substrate, but with all atoms fixed in the precise geometries they adopt in the 
adsorption structure. 
Notwithstanding the fact that any energy decomposition scheme is 
to some extent arbitrary \cite{michaelides04}, 
the estimate of the water-metal bonding, $E_{\rm ads}^{\rm wm}$,
is simply taken  as the difference between $E_{\rm ads}$ and
$E_{\rm gas}^{\rm ww}$. 
We have also examined the non-local correlation
part, $E^{\rm nlc}$, of the exchange-correlation energy
which is obtained directly from the calculation. We note that $E^{\rm nlc}$
includes a part which acts as a semi-local correction to LDA correlation, i.e. in 
a similar spirit as PBE correlation works. Thus the ``non-local" interaction means 
interaction beyond LDA correlation.
The corresponding 
attraction due to the non-local correlation in  the adsorption energy per water molecule
can be calculated as:

\begin{equation}\label{adenenlc}
	E_{\rm ads}^{\rm nlc}= \frac{E^{\rm nlc}[{\rm H_2O/M}] - E^{\rm nlc}[{\rm M}]- nE^{\rm nlc}[{\rm H_2O}]}{n},
\end{equation}

\noindent an analogous expression to Eq. \ref{adene}, but where total energies are
substituted by their non-local correlation part.

\section{Results and discussion}\label{sec:res}

\subsection{Adsorbed water monomers}

First we consider the adsorption of an isolated water monomer
on a series of metal surfaces. This is the simplest water adsorption system possible
to investigate the role of vdW interactions with the metal, since no H-bonds
between water molecules are present. In the first place we explored the effect of
vdW interactions in determining the most stable adsorption site on a selection
of close-packed metal surfaces. Considering the most stable adsorption site, we
then extended the study to a larger number of substrates to investigate in a
systematic manner the dependence of vdW interactions with the nature of the
metal.

\begin{figure}[htb]
\includegraphics[width=0.85 \textwidth]{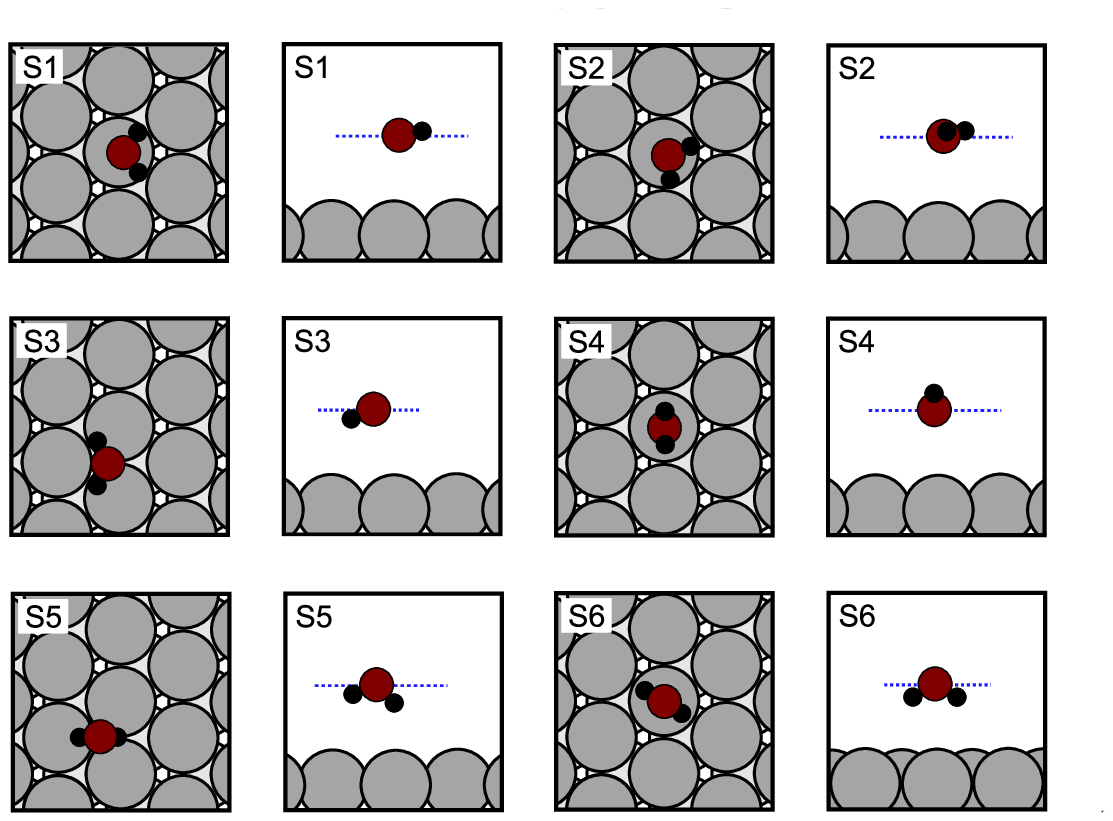}
\caption[]{Top and side views of water monomers adsorbed at different
sites on a close-packed metal surface. Small black, red, and grey
spheres stand for H, O, and metal atoms, respectively.
\label{fig:sites}}
\end{figure}

Previous DFT studies with semi-local functionals have predicted that
water adsorbs on atop sites on
close-packed surfaces \cite{michaelides03,meng04,knudsen07,carrasco09b}.
Since recent studies have shown that vdW forces can alter adsorption
structures \cite{atodiresei09,mittendorfer11,liu2012b,li12}, it is important to establish if vdW
changes the preferred site for water adsorption on metals. 
To this end we computed the optB88-vdW  (and for reference PBE)
adsorption energies of a water monomer on Ag(111), Au(111), and Ru(0001).
More than 30 different adsorption sites and molecule orientations were explored, 
from which 6 stable and representative structures are shown in Fig. \ref{fig:sites}. We found that  the
most stable adsorption site with the optB88-vdW functional is the atop site, consistent with
previous work and our own new PBE calculations reported in Table \ref{tab:sites}. These results
indicate that accounting for vdW interactions does not lead to a qualitative change in the 
mechanism governing the water-metal adsorption geometry of adsorbed
monomers \cite{carrasco09b}. Nevertheless, vdW interactions increase 
the strength of the bond with the substrate. In particular, on the noble metals,
orientations of the water molecule with the oxygen atom away from the surface
(S5 and S6) have almost no binding to the surface at the PBE level, whereas
optB88-vdW adsorption energies
are relatively large (Table \ref{tab:sites}). 
This will very likely be of relevance to water monomer
diffusion, with the diffusing water molecule able to access many more stable configurations
when dispersion forces are accounted for.

\begin{figure}[htb]
\includegraphics[width=0.99 \textwidth]{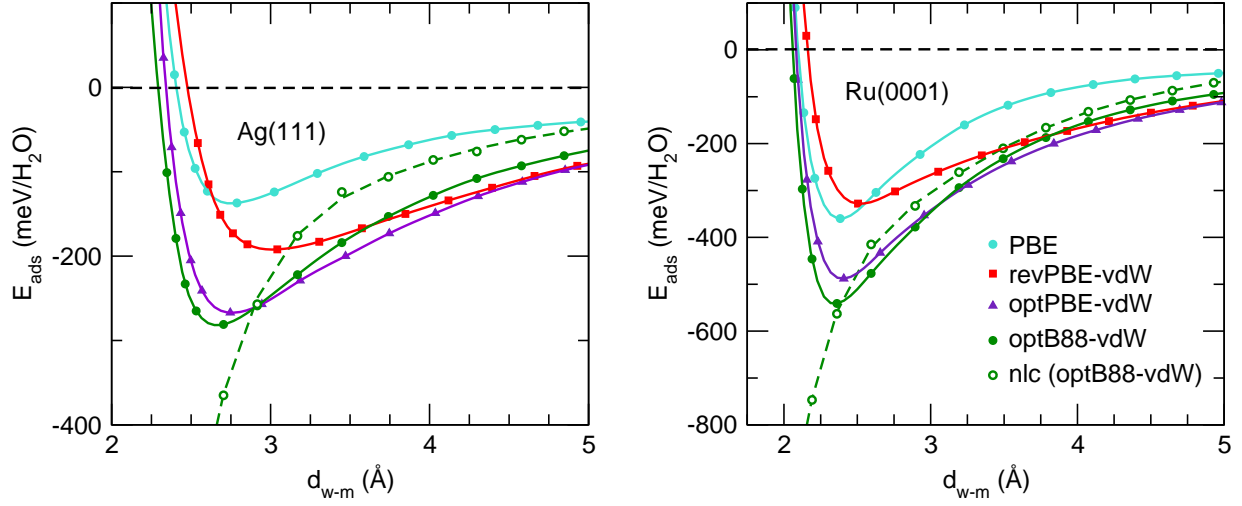}
\caption[]{Adsorption energy of H$_2$O monomer on Ag(111) and Ru(0001)
as a function of the vertical water-metal distance ($d_{\rm w-m}$), defined as
the distance between the O atom
of a water molecule and the nearest metal atom on the surface. Four
different  
functionals are considered and the optB88-vdW non-local correlation (nlc)
contribution to the total energy (Eq. \ref{adenenlc}) is also shown (open circles and dashed line).
The lines are merely a guide to the eye.
\label{fig:b_curve_plot}}
\end{figure}

In order to gain more insight into the role of  vdW forces on the adsorption 
energy and their dependence on the specific 
functional chosen, we calculated the 
PBE, revPBE-vdW, optPBE-vdW, and optB88-vdW binding curves of a water
monomer on Ag(111) and Ru(0001) as a function of the
distance between oxygen and the metal atom underneath (Fig. \ref{fig:b_curve_plot}). 
Each point in the graph is obtained by keeping the $z$ coordinate of
the water oxygen atom fixed while relaxing the rest of 
the atoms in the system except 
the 3 bottom layers of the metal slab. Of the functionals considered, PBE predicts the 
weakest interaction at  long-range. 
Although revPBE-vdW recovers the long-range attraction, the 
interaction strength around the equilibrium distance is only slightly larger than that 
obtained from PBE on Ag and on Ru the revPBE-vdW binding minimum is 
actually shallower than that obtained with PBE.
At long-range,  the energies of the 
optPBE-vdW and optB88-vdW functionals are similar to revPBE-vdW, but
on approaching the surface become significantly more attractive. 
Computing the non-local correlation contribution to the total adsorption 
energy reveals that this is indeed the leading attractive term between the water
molecule and the surface as observed 
previously by Hamada and co-workers \cite{hamada10}.

Considering the most stable adsorption site and adsorbed water orientation identified 
previously (S1), we examine now the adsorption energy of water monomers on
a large range of metal substrates: Al(111), Cu(111), Cu(110), Ru(0001), Rh(111),
Pd(111), Ag(111), Pt(111), and unreconstructed Au(111). Also included in this
systematic study are results from the original revPBE-vdW functional for comparative purposes
with respect to optB88-vdW. The computed adsorption energies and optimized distances are
summarized in Table \ref{tab:monomer}. 
A key observation is that revPBE-vdW does not always enhance
adsorption energies with respect to PBE. Indeed the revPBE-vdW adsorption energies can 
either be larger or smaller than PBE. 
This is consistent with previous studies for various adsorption systems where the revPBE-vdW adsorption energies can either
be similar to or slightly smaller than PBE \cite{dion04,puzder06,ziambaras07,lee10,hamada10b,klimes10,mittendorfer11,liu2012}.  
This behaviour is a direct 
consequence of the underlying overly repulsive revPBE exchange in the revPBE-vdW kernel, which 
also causes pure revPBE adsorption energies to be underestimated compared to PBE. 
Indeed water-metal
distances ($d_{\rm w-m}$)---defined here as the distance between the
O atom of a water molecule and the nearest metal atom on the surface---optimized with
revPBE-vdW are substantially larger (up to 0.32 \AA) than those obtained from PBE.
Another key observation is that
optB88-vdW consistently provides larger adsorption energies than PBE and similar
water-metal distances (within 0.1 \AA) for all investigated systems.
As observed before in the binding curves of Ag(111) and Ru(0001),
we see that non-local correlation ($E_{ads}^{\rm nlc}$)
is again the principal attractive contribution to the total adsorption energy ($E_{ads}$) as 
shown in Table \ref{tab:monomer}. Specifically, $E_{ads}^{\rm nlc}$ is in most cases
considerably larger than $E_{ads}$ at the equilibrium geometry independent of the metal
and functionals considered.
Although the contribution of vdW dispersion
forces to the non-local correlation at equilibrium distances is probably not dominant, it clearly enhances
the total adsorption
energy as indicated by the fact that typical optB88-vdW adsorption energies are in
general 110-190 meV/H$_2$O larger than PBE values (Table \ref{tab:monomer}).
Interestingly, larger adsorption energy enhancements are found on more reactive metals
than on noble ones. For example, when comparing Pd(111) and Ag(111), the difference 
between PBE and optB88-vdW adsorption energies is 180 meV/H$_2$O and 
144 meV/H$_2$O, respectively. Similarly, the vdW enhancement is also larger when
comparing the adsorption on Pt(111) and Au(111): 186 meV/H$_2$O and 
158 meV/H$_2$O, respectively. 
This behavior can be rationalized in terms
of the equilibrium water-metal distance in each case as discussed recently by 
Liu {et al.} for benzene adsorption on metal surfaces
\cite{liu2012}. As shown in Table \ref{tab:monomer}
the water molecule adsorbs closer to the surface on reactive metals such as 
Pd and Pt than it does on more noble substrates such as Ag and 
Au. 
The shorter water-metal distances lead to larger vdW dispersion forces and,
therefore, larger total adsorption energies.

Overall we have shown that optB88-vdW (and for the systems considered optPBE-vdW)
increases adsorption energies by accounting for vdW interactions without substantially
altering adsorption sites or bonding distances with respect to PBE. In contrast, although
revPBE-vdW is able to catch similarly vdW interactions at long distances, adsorption
energies are not increased beyond those obtained with PBE. Unfortunately, experimental
data is not yet available against which we can benchmark our computed adsorption
energies. This is mainly because unlike simpler adsorption systems such as CO or noble
gas adsorption, water molecules readily diffuse and cluster on the surface, forming
complex cluster and overlayer structures (e.g. the clusters discussed in
section \ref{sec:clusters}). Clusters and overlayers obviously complicate the analysis of 
experimental adsorption enthalpy determinations, especially with approaches such as
temperature programmed desorption~\cite{karlberg06}. Although experimental adsorption
energies for water monomers on well-defined metal surfaces are lacking, we note that in
a recent study it was possible to perform the first single crystal adsorption microcalorimetry
measurements of the energy of a well-deÞned water adstructure (a partially dissociated
water adlayer stabilized by hydrogen bonding)~\cite{lew11}. For this one overlayer
structure we were able to show that the inclusion of vdW forces using optB88-vdW was
crucial for achieving quantitative agreement with experiment \cite{lew11}; a similar conclusion
has also been drawn for the adsorption of benzene on Au(111) and Pt(111)~\cite{liu2012}.
However, in general, accurate experimental measurements of adsorption energies of water
on well-defined surfaces are severely lacking and urgently
needed \cite{carrasco12,klimes12}.

\subsection{Adsorbed water clusters}
\label{sec:clusters}

Let us focus now on more complex water adsorption systems than simple 
adsorbed water monomers, in which we simultaneously consider the interplay of
water-metal and water-water interactions. To this end we have considered a
series of adsorbed water clusters with between 2 and 6 water 
molecules on Cu(110). We have considered Cu(110) because of a number
of recent STM experiments for low coverages of water on this surface
\cite{carrasco09,kumagai11,yamada06,lee08,forster11}
and because the open (110) surface allows us 
to explore a rich variety of isomers for a given water cluster and,
in so doing, analyze the role of vdW in determining the relative energies of various 
adsorbed water clusters.

\begin{figure}[htb]
\includegraphics[width=0.90 \textwidth]{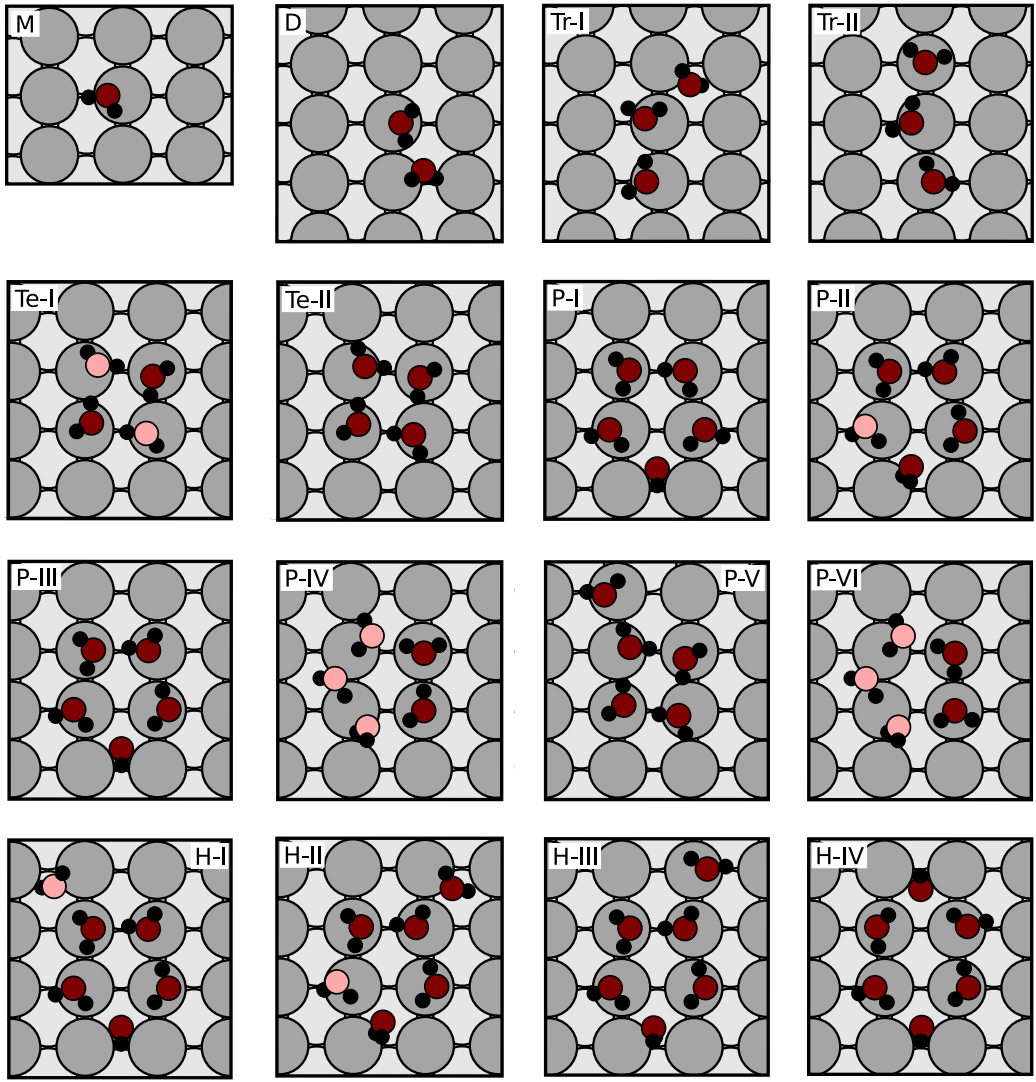}
\caption[]{Top view of water monomer (M), dimer (D), trimers (Tr-I and Tr-II),
tetramers (Te-I and Te-II), pentamers (P-I, P-II, P-III, P-IV, P-V, and P-VI), and
hexamers (H-I, H-II, H-III, and H-IV) adsorbed on Cu(110). Small black, red,
dark grey, and light grey spheres stand for H, O, and Cu in the first and second
layer, respectively. Light red spheres indicate water molecules that are 
relatively far 
away from the metal surface (typically $>$3 \AA).
\label{fig:cluster_models}}
\end{figure}

For each of the clusters examined we considered a range of different
initial geometries which upon optimization lead to a number of stable or
metastable water clusters, as shown in Fig. \ref{fig:cluster_models}. 
In Table \ref{tab:clusters1}
we summarize
the adsorption energies of all
these structures employing various functionals. For the adsorbed trimers and tetramers 
our results are in general agreement with a recent study by
Kumagai {\it et al.} \cite{kumagai11}, who by combining STM and DFT identified
chain trimers and cyclic tetramers as the most stable species.

Consistent with the results of the water monomers we find for the clusters that
including vdW interactions through optB88-vdW leads to a substantial
increase ($>$120 meV/H$_2$O) of the total adsorption energy
when compared with
PBE. The optPBE-vdW adsorption energies are typically smaller than those from
optB88-vdW by $\approx$50 meV/H$_2$O and revPBE-vdW  values
are in all cases less than PBE by at least 45 meV/H$_2$O.
Although in most situations the structures obtained with PBE and optB88-vdW
are similar, there are some occasions when accounting for van der Waals
dispersion forces can alter the relative stabilities of the clusters considered.
In particular, optB88-vdW tends to favor planar structures over buckled
ones. For example, in the case of water tetramers the buckled cluster (Te-I) is 
preferred at the PBE or even revPBE-vdW level, over the planar tetramer (Te-II) by
about 10--15 meV/H$_2$O.
However, the situation is reversed when considering
optB88-vdW which predicts the planar tetramer (Te-II) to be 15 meV/H$_2$O more 
stable than the buckled one. It turns out that this result can explain the absence
of corrugation in the STM images of Kumagai {\it et al.} \cite{kumagai11}.
The lack of buckling 
was attributed to tip-induced reorientation or fast dynamical 
fluctuations between the two isomers \cite{kumagai11}.
However, here we see that there is no
need to invoke such arguments as the most stable structure when vdW is
accounted for is planar.
Similarly, the preference for planar adstructures when switching on vdW dispersion
through optB88-vdW is again observed in the case of the P-II and P-III pentamers,
and the H-II and H-III hexamers (Table \ref{tab:clusters1}).

\begin{figure}[htb]
\includegraphics[width=0.65 \textwidth]{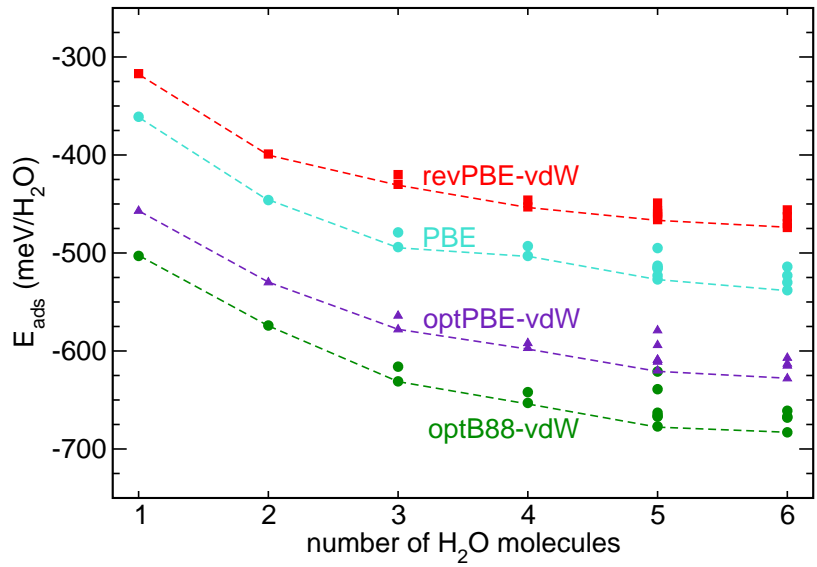}
\caption[]{Adsorption energies of different sized water clusters adsorbed on Cu(110)
using PBE, revPBE-vdW, optPBE-vdW, and optB88-vdW. Dashed lines connect 
the most stable isomers for a given number of H$_2$O molecules with each functional.
\label{fig:size}}
\end{figure}

Another important observation is that the clusters become more stable as the 
clusters increase in size (Fig. \ref{fig:size}). This is consistent with what has been
observed on other surfaces \cite{michaelides07,meng04} and gas phase clusters
\cite{liu96, santra07} and is due to 
the formation of H-bonds
between water molecules and to a well-known cooperative effect,
i.e., the increase of the number of H-bonds within the clusters results in stronger
H-bonds.
Interestingly, the nature of this cooperative effect shows a very small 
dependence with respect to the particular functional considered: both PBE and 
the vdW-DFs present similar trends, only the magnitude of the total 
adsorption energy is shifted  (Fig. \ref{fig:size}).
We see, therefore, that vdW forces seem to
play a minor role in describing cooperative water-water effects in these
systems.

In order to gain deeper insight into these adsorption systems, we
decomposed the total adsorption energy into water-water and 
water-metal bonding contributions as given by Eq. \ref{eq:ww}.
As shown in Table \ref{tab:clusters1},
the water-water bonding contribution
to the adsorption energy
obtained from the various functionals is very similar:
the PBE, optPBE-vdW, 
and optB88-vdW results are all within 13 meV/H$_2$O for all adsorption
systems (the revPBE-vdW water-water interactions are systematically smaller in 
all cases). It is clear, therefore, that the main 
effect of vdW forces on the total interaction comes from the water-metal
bonding. Indeed the approximately 120 meV/H$_2$O increase in the 
adsorption energy comes almost exclusively from the water-metal bonding.
This is consistent with our earlier work on ice-like films on Ru and Cu
\cite{carrasco11}.  
Unlike the GGA functionals where the contribution to the binding comes from
density overlap, the non-local functionals give binding even between non-overlapping
electron densities \cite{lazic12}. 
Therefore, when considering non-local functionals not only a small overlap region gives
contribution to the adsorption energy, but a larger part of the surface too. This leads to
a stronger water-metal interaction for non-local functionals than GGA functionals.
Moreover, this also explains that the shift in the adsorption energy caused
by the vdW density functional with respect to PBE is almost constant between different
clusters.
In general, water-metal bonding is greater than 
water-water bonding
and is responsible for
the relative stability of the isomers and preference for more planar
rather than buckled isomers, independent of the considered
functional. 
Upon comparing the strength of the water-water interaction for the 
various clusters considered, we find that the water-water bonding is strongest 
in the water tetramers (Te-I and Te-II).
Larger clusters have
larger total adsorption energies, but their water-water bonding is substantially 
reduced with respect to the optimal H-bond configuration offered by 
a tetragonal arrangement on the Cu(110) surface.
This becomes especially evident in
the case of pentamers, where the most stable isomer, P-I, shows 
actually a relatively small optB88-vdW water-water bonding (--169 meV/H$_2$O)
when compared for  example to isomer P-V (--260 meV/H$_2$O).

\begin{figure}[htb]
\includegraphics[width=0.95 \textwidth]{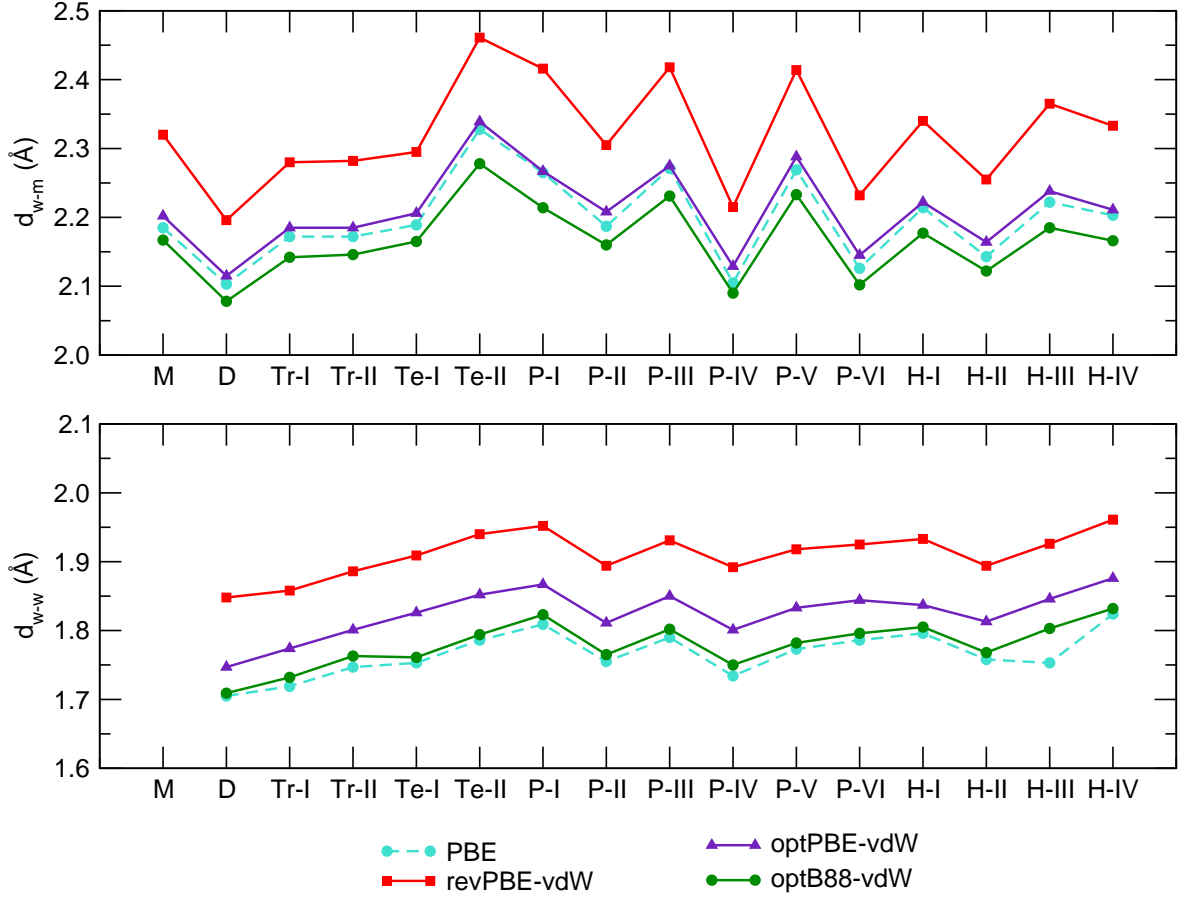}
\caption[]{Averaged nearest neighbor water-metal, $d_{\rm w-m}$ (top), and water-water,
$d_{\rm w-w}$ (bottom), distances for the water clusters depicted in
Fig. \ref{fig:cluster_models} using PBE, revPBE-vdW, optPBE-vdW, and optB88-vdW. 
The lines connecting the points are there to guide the eye.
\label{fig:distances_plot}}
\end{figure}

Regarding the effect of vdW forces on the adsorption geometry of the 
clusters, we summarize in Table \ref{tab:clusters2} and
Fig. \ref{fig:distances_plot} the averaged water-metal and water-water
distances computed with different functionals. Essentially PBE geometries
do not differ significantly with respect to the optB88-vdW
or optPBE-vdW
functionals for both water-metal and water-water
distances. All distances are within 0.05 \AA\ of PBE for optB88-vdW
and within 0.09 \AA\ of PBE for optPBE-vdW.
On the contrary, revPBE-vdW predicts larger 
water-metal ($<$0.15 \AA) and water-water ($<$0.14 \AA) distances. It is
interesting, therefore, that despite a noticeable enhancement in water-metal 
bonding when vdW is accounted for, the water-metal distances remain rather
similar. 

\section{Conclusion}\label{sec:conclusion}

We have investigated the role of vdW dispersion forces in water-metal bonding
by considering different non-local vdW-DFs and a range of water monomers and
small water clusters on a series of metal surfaces. Analysis of our results reveals
an enhancement 
of adsorption energies (typically $>$110 meV/H$_2$O) due to vdW interactions
with respect to the widely used PBE functional.
The increase in the adsorption energy comes almost exclusively from an
increased water-metal interaction, with the water-water interaction within
the adsorbed clusters being essentially unaffected by the inclusion of vdW
dispersion forces.
As a consequence, we
observe that in general the 
explicit consideration of vdW dispersion forces does not alter the relative stabilities
of structures 
predicted by PBE. In addition, despite increases in the total
adsorption energies, the adsorption sites remain unchanged and adsorption 
geometries (water-metal bond and water H-bond lengths) are very similar when
comparing PBE and optB88-vdW in general. 
On the few occasions when the 
PBE and optB88-vdW structures differ, the optB88-vdW structures are 
flatter with any high-lying water molecules brought closer to the surface, 
which for the particular case of the water tetramer on Cu(110) produces an 
adsorption structure in better agreement with experiment.
This is
a direct consequence of the fact that non-local correlation 
is enhanced by shorter water-metal distances.
It is interesting to note that the general similarity between the structures obtained with and 
without vdW dispersion forces is in contrast to what has been found for gas
phase water clusters. 
In particular for the water hexamers---where this issue
has been considered in greatest detail---the relative energies of the relevant 
low energy isomer structures differ completely depending on whether vdW
is accounted for or not \cite{santra07}.

\section{Acknowledgements}

JC is a Ram\'{o}n y Cajal fellow and Newton Alumnus supported by the Spanish
Government and The Royal Society, respectively.
AM is supported by the European Research Council and the Royal Society
through a Royal Society Wolfson Research Merit Award.
We are grateful for computer time to UCL
Research Computing, the London Centre for Nanotechnology, and the UK's national high performance computing service 
HECToR (from which access was obtained via the UK's Material Chemistry
Consortium,  EP/F067496). 


\newpage

\newpage

\newpage


\newpage

\newpage


\newpage

\newpage

\newpage

\begin{table}[tbp]
\centering
\caption{\label{tab:sites} Adsorption energies (in meV/H$_2$O) of an isolated water 
monomer on three different metal surfaces with PBE and the optB88-vdW functionals.
Six different adsorption sites
and adsorbed water geometries are shown (see Fig. \ref{fig:sites}). 
S1 is the most stable adsorption structure on all surfaces with all functionals.
Negative adsorption energies correspond to favorable (exothermic) adsorption.
}
\begin{ruledtabular}
\begin{tabular}{lcccccc}
                     &     S1   &     S2     &     S3     &     S4 & S5 & S6     \\               
\hline
PBE              &  & & & & &\\
\hline                                                
Ag(111)         & --137 & --134 & --41  &   --1    & 3      & 25\\
Au(111)         & --123 & --122 & --68  & 30      & 8      & 62\\
Ru(0001)       & --360 & --356 & --121 & --130 & --17 & --17\\   
\hline
optB88-vdW  &  & & & & &\\
\hline                                                
Ag(111)         &  --281 & --280 & --195 & --151 & --117 & --103\\ 
Au(111)         &  --281 & --281 & --213 & --120 & --122 & --76\\
Ru(0001)      &  --541 & --538 & --271 & --334 & --131 & --125\\                                                                                          
\end{tabular}
\end{ruledtabular}
\end{table}

\newpage

\begin{table}[tbp]
\centering
\caption{\label{tab:monomer} Adsorption
energies ($E_{\rm ads}$ in meV/H$_2$O) and the optimized distance between the O atom
of a water molecule and the nearest metal atom on the surface ($d_{\rm w-m}$ in \AA) 
for water monomers at the equilibrium adsorption site
(S1 in Fig. \ref{fig:sites}) on all the metal surfaces investigated. Results for the PBE, revPBE-vdW, and optB88-vdW functionals are reported.
}
\begin{ruledtabular}
\begin{tabular}{lccccccc}
       Metal & PBE & \multicolumn{2}{c}{revPBE-vdW} & \multicolumn{2}{c}{optPBE-vdW} & \multicolumn{2}{c}{optB88-vdW }\\
                   & $E_{\rm ads}$   
                   & $E_{\rm ads}$ & $E_{\rm ads}^{\rm nlc}$ 
                   & $E_{\rm ads}$ & $E_{\rm ads}^{\rm nlc}$ 
                   & $E_{\rm ads}$ & $E_{\rm ads}^{\rm nlc}$ \\
\hline     
Al(111)      & --192            &   --165   & --380 &    --279   & --436 & --302 & --476 \\ 
Cu(111)    & --167              &  --201   & --360 &    --292    & --438 & --319    & --456\\
Ru(0001)  & --360              & --328    & --490 &    --488   & --530 & --541    & --563 \\  
Rh(111)    &  --306             & --293     & --466 &    --442  & --527 & --497    & --579 \\
Pd(111)    & --243              & --263     & --423 &    --380   & --458 & --423   & --529    \\ 
Ag(111)    & --137              & --192      & --260 &    --267  & --345 & --281   &  --365  \\
Pt(111)     & --217              & --241      & --391 &   --358  & --495 & --403    & --500  \\
Au(111)    & --123             & --192        & --392 &    --264  & --398 & --281   &  --433  \\
Cu(110)    & --361              & --317      & --389&     --457  & --486 & --503 &  --474   \\
\hline
\multicolumn{8}{l}{$d_{\rm w-m}$}  \\
\hline                 
Al(111)       & 2.232   & \multicolumn{2}{c}{2.503}  &   \multicolumn{2}{c}{2.248} &  \multicolumn{2}{c}{2.195} \\
Cu(111)      & 2.507    & \multicolumn{2}{c}{2.795}  &  \multicolumn{2}{c}{2.495}  & \multicolumn{2}{c}{2.422}\\
Ru(0001)    & 2.383    & \multicolumn{2}{c}{2.505} &  \multicolumn{2}{c}{2.408}   & \multicolumn{2}{c}{2.361}\\
Rh(111)      & 2.359   & \multicolumn{2}{c}{2.519}   &  \multicolumn{2}{c}{2.389}  & \multicolumn{2}{c}{2.334}\\
Pd(111)       & 2.469    & \multicolumn{2}{c}{2.714}   &  \multicolumn{2}{c}{2.503}  & \multicolumn{2}{c}{2.419}\\
Ag(111)       & 2.788   & \multicolumn{2}{c}{3.042}   &  \multicolumn{2}{c}{2.746}  & \multicolumn{2}{c}{2.704}\\
Pt(111)        & 2.499   & \multicolumn{2}{c}{2.815}    &  \multicolumn{2}{c}{2.555}   & \multicolumn{2}{c}{2.498}\\
Au(111)      & 2.836   & \multicolumn{2}{c}{3.084}    &    \multicolumn{2}{c}{2.853}  & \multicolumn{2}{c}{2.750}\\
Cu(110)      & 2.185    & \multicolumn{2}{c}{2.320}   &   \multicolumn{2}{c}{2.202}  & \multicolumn{2}{c}{2.167}\\
\end{tabular}
\end{ruledtabular}
\end{table}

\newpage

\begin{table}[tbp]
\centering
\caption{\label{tab:clusters1} Computed adsorption energies
($E_{\rm ads}$) of water monomer (M),
dimer (D), trimers (Tr), tetramers (Te), pentamers (P), and hexamers (H) adsorbed on 
Cu(110) using PBE, revPBE-vdW, optPBE-vdW, and optB88-vdW DFs. Water-water
($E_{\rm gas}^{\rm ww}$) and water-metal ($E_{\rm ads}^{\rm wm}$)
bonding contributions are also given. All values are in meV/H$_2$O.
}
\begin{ruledtabular}
\begin{tabular}{c|ccc|ccc|ccc|ccc}

                 & \multicolumn{3}{c|}{PBE} & \multicolumn{3}{c|}{revPBE-vdW} & \multicolumn{3}{c|}{optPBE-vdW} & \multicolumn{3}{c}{optB88-vdW}\\
                  &  $E_{\rm ads} $ & $E_{\rm gas}^{\rm ww}$ & $E_{\rm ads}^{\rm wm}$
                  & $E_{\rm ads} $ & $E_{\rm gas}^{\rm ww}$ & $E_{\rm ads}^{\rm wm}$
                  & $E_{\rm ads} $ & $E_{\rm gas}^{\rm ww}$ & $E_{\rm ads}^{\rm wm}$
                  & $E_{\rm ads} $ & $E_{\rm gas}^{\rm ww}$ & $E_{\rm ads}^{\rm wm}$ \\
 \hline
 M       & --361 & --      & --361 & --317 & -- & --317 & --457 & -- & --457 & --503  & -- & --457 \\ 
 \hline
 D       &   --446 & --94 & --352 & --399 & --82 & --317 &  --530 & --92 & -438 & --574 & --89 & --485 \\
 \hline
Tr-I     & --494 & --130 & --364 & --430 & --110 & --320 & --578 & --128 & --450 & --631& --126 & --505\\
 Tr-II   & --479 & --137 & --342& --420 & --113 & --307 & --564 & --133 & --431 & --616 & --133 & --483\\
 \hline
 Te-I   & --503 & --304 & --199 & --453 & --244 & --209 & --592 & --291 & --301 & --642 & --304 & --338\\
 Te-II  & --493 & --282 & --211 & --446 & --225 & --221 & --597 & --272 & --325 & --653 & --282 & --371\\
 \hline
 P-I     & --527 & --168 & --359 & --466 & --150 & --316 & --620 & --174 & --446 & --677  & --169 & --508\\
 P-II    & --523 & --206 & --317 & --463 & --176 & --287 & --609 & --207 & --402 & --663  & --205 & --458\\
 P-III  & --516 & --209 & --307 & --459  & --178 & --281 & --611 & --209 & --402 & --666  & --208 & --458\\
P-IV  & --515 & --195 & --320 & --460  & --172 & --288 & --594 & --200 & --394 & --639 & --197 & --442\\
 P-V   & --513 & --260 & --253 & --457 & --212 & --245 & --609 & --253 & --356 & --667 & --260 & -407\\
 P-VI  & --495 & --158 & --337 & --449 & --142 & --307 & --579 & --166 & --413 & --621 & --159 & --462\\
 \hline
 H-I    & --538 & --159 & --379 & --474 & --142 & --332 & --628 & --163 & --465 & --683 & --159 & --524\\
 H-II   & --530 & --190 & --339 & --470 & --165 & --305 & --615 & --193 & --421 & --667 & --190 & --477\\
 H-III  & --523 & --193 & --330 & --463 & --166 & --297 & --613 & --195 & --418 & --668 & --192 & --476\\
 H-IV  & --514 & --156 & --358 & --456 & --141 & --315 & --607 & --163 & --444 & --661 & --155 & --506\\
\end{tabular}
\end{ruledtabular}
\end{table}

\newpage

\begin{table}[tbp]
\centering
\caption{\label{tab:clusters2} Averaged nearest neighbor water-metal ($d_{\rm w-m}$)
and H-bond ($d_{\rm w-w}$) distances for different water clusters adsorbed on Cu(110)
using  PBE, revPBE-vdW, optPBE-vdW, and optB88-vdW DFs. All values are in \AA.
}
\begin{ruledtabular}
\begin{tabular}{c|cccccccc}
&  \multicolumn{2}{c}{PBE}  &  \multicolumn{2}{c}{revPBE-vdW}  & \multicolumn{2}{c}{optPBE-vdW} &  \multicolumn{2}{c}{optB88-vdW} \\
& $d_{\rm w-m}$ & $d_{\rm w-w}$ &  $d_{\rm w-m}$ & $d_{\rm w-w}$ &  $d_{\rm w-m}$ & $d_{\rm w-w}$ &  $d_{\rm w-m}$ & $d_{\rm w-w}$ \\
 \hline
M       &   2.185 & --          &   2.320 & --        &   2.202 & --         &   2.167 & --  \\ 
D       &   2.103  & 1.705 &   2.196 & 1.848 &   2.115 & 1.747 &    2.078 & 1.709  \\
Tr-I    &   2.172  & 1.719 &   2.280  &1.858 &   2.185 & 1.774 &     2.142 & 1.732  \\
Tr-II   &   2.172  & 1.747 &   2.282  &1.886 &   2.185 & 1.801 &    2.146 & 1.763  \\
Te-I   &   2.189  & 1.753 &   2.295  &1.909 &   2.206 & 1.826 &     2.165 & 1.761  \\
Te-II  &   2.328  & 1.786 &   2.461  &1.940 &   2.339 & 1.852 &     2.278 & 1.794  \\
P-I     &   2.265  & 1.809 &   2.416  &1.952 &   2.267 & 1.867 &     2.214 & 1.823  \\
P-II    &   2.187  & 1.755 &   2.305  &1.894 &   2.208 & 1.811 &     2.160 & 1.765  \\
P-III   &   2.271  & 1.790 &   2.418  &1.931 &   2.275 & 1.850 &     2.231 & 1.802  \\
P-IV  &   2.105  & 1.734 &   2.215  & 1.892 &   2.129 & 1.801 &    2.090 & 1.750  \\
P-V   &   2.269  & 1.773 &   2.414  & 1.918 &   2.288 & 1.833 &   2.233 & 1.782  \\
P-VI  &   2.126  & 1.786 &   2.232  & 1.925 &   2.145 & 1.844 &     2.102 & 1.796  \\
H-I    &   2.214  & 1.796 &   2.340  & 1.933 &   2.222 & 1.837 &    2.177 & 1.805  \\
H-II   &   2.143  & 1.758 &   2.255  & 1.894 &   2.164 & 1.813 &     2.122 & 1.768  \\
H-III  &   2.222  & 1.753 &   2.365 & 1.926 &    2.238 & 1.846 &     2.185 & 1.803   \\
H-IV  &  2.203  & 1.824 &   2.333 & 1.961 &    2.211 & 1.876 &      2.166 & 1.832 \\
\end{tabular}
\end{ruledtabular}
\end{table}

\newpage

\end{document}